# Complex impedance measurements of calorimeters and bolometers: Correction for stray impedances

Mark. A. Lindeman, Kathleen A. Barger, Donald E. Brandl, S. Gwynne Crowder, Lindsay Rocks, and Dan McCammon
*University of Wisconsin, Madison, Wisconsin 53706*

Impedance measurements provide a useful probe of the physics of bolometers and calorimeters. We describe a method for measuring the complex impedance of these devices. In previous work, stray impedances and readout electronics of the measurement apparatus have resulted in artifacts in the impedance data. The technique allows experimenters to find an independent Thevenin or Norton equivalent circuit for each frequency. This method allows experimenters to easily isolate the device impedance from the effects of parasitic impedances and frequency dependent gains in amplifiers. © *2007 American Institute of Physics*. [DOI: 10.1063/1.2723066]

## I. INTRODUCTION

The measurement of a complex impedance is a becoming a standard tool for characterizing electronic and thermal properties of microcalorimeters and bolometers.[1–6] To measure the complex impedance, we add a small ac perturbation, either white noise or a swept sine wave, to the dc bias of the thermistor. We measure the response of the device as a function of frequency $f$. The perturbation is kept small so that the response of the device to the perturbation is linear. When no dc bias is applied to a thermistor, it responds to the ac perturbations as an ordinary ohmic resistor does.

For the purpose of this article, we are primarily interested in the component of the thermistor's impedance that provides information about the thermal physics of the calorimeter. This impedance $Z$ is associated with dissipation in the thermistor. It does not include parasitic impedances associated with capacitance or inductance of the thermistor. When the thermistor is unbiased, the impedance $Z$ is frequency independent, real valued, and equal to the thermistor resistance. When a thermistor is biased, additional factors affect the impedance $Z$ of the thermistor, making it a complex valued function of frequency $f$. Under bias, the impedance $Z(f)$ depends on how resistance varies with temperature and with current or voltage. The impedance also depends on the thermal physics of the various components of a calorimeter or bolometer, including the thermal couplings between thermistor components and the refrigerator, heat capacitances of the components, and electrothermal feedback. Models for the impedance $Z(f)$ of various kinds of calorimeters and bolometers have been presented by a number of authors.[1,7,8]

We measure transfer functions to obtain the impedance of a microcalorimeter. In the case of our Si thermistors, the transfer functions are proportional to the voltage $V(f)$ across the thermistor divided by the ac bias $V_{ac}(f)$ applied the bias circuit. The transfer function is also proportional to the gain associated with amplifiers and electronic filters. An example circuit is shown in Fig. 1. In measuring *unbiased transfer functions*, no dc bias is applied to the thermistor, therefore it behaves as an ordinary resistance $R$. Unbiased transfer functions are useful for characterizing the bias circuit, because the thermistor impedance is known to be frequency independent. (Any other situation in which the thermistor has known impedance is also useful for characterizing the bias circuit.) *Biased transfer functions*, in which the thermistor is biased, are used to measure the complex impedance $Z(f)$ of the thermistor. When thermistors are biased, thermal effects can cause $Z(f)$ to be very different than resistance $R$.

A transfer function depends not only on $Z(f)$, but also on the parasitic impedance of the thermistor, on gains associated with the bias and readout electronics, and on various impedances in the circuits used to bias and readout the devices. These complexities make it difficult to compute $Z(f)$ from a measured transfer function. If an unexpected feature is seen in the data, it is difficult to know if it originates in the thermistor or in the bias or readout electronics.

The effects of stray impedance can be accounted for by modeling the bias circuit as was done by Lindeman[1] or Vaillancourt.[2] An example circuit model is illustrated in Fig. 1. This model includes resistors and the thermistor impedance plus a number of capacitors to represent the stray capacitances in the bias circuit and of the thermistor itself. Vaillancourt *et al.* pointed out that effects of stray impedances on impedance measurements could be mitigated by fitting the circuit model to an unbiased transfer function to determine the values of the impedances in the circuit model. An example of a best fit of this circuit model to one of our unbiased transfer functions is shown in Fig. 2. Once the strays are determined from the fit, the circuit model is then used with a biased transfer function to solve for the impedance $Z(f)$ of the thermistor. However, the accuracy of this approach depends on how well the bias circuit model matches the actual circuit. Having precise measurements of amplifier gains over the range of useful frequencies is important also. In practice, we find the bias circuit model fails at frequencies above several kHz. The model produces artifacts in $Z(f)$ possibly because we do not accurately model distributed impedances in the actual circuit or the actual behavior of our amplifiers. Even if the model of the bias circuit fits transfer function data well, there is no guarantee that the circuit model is accurate, as we will describe below. Therefore, this method can result in significant experimental errors in the measurement of $Z(f)$.

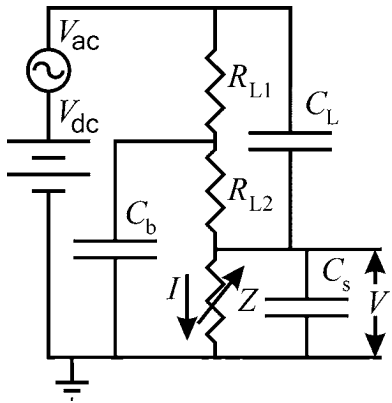

FIG. 1. A bias circuit model. A dc bias $V_{dc}$ plus a small ac perturbation $V_{ac}$ is applied through the load resistance of 172 M$\Omega$ to the thermistor. The thermistor is represented by the variable impedance $Z$ and the stray capacitance $C_s$. The capacitors $C_b$ and $C_L$ represent stray capacitance in the physical bias circuit. The resistors $R_{L1}$ and $R_{L2}$ each represent half of the load resistance. A transfer function measures the ac component of the voltage $V$ across the thermistor divided by ac bias $V_{ac}$. If the illustrated circuit model is correct, we can estimate the capacitances by fitting this model to transfer function data plotted in Fig. 2. The thermistor capacitance $C_s$ is found to be 30 pF. The capacitors $C_b$ and $C_L$ are 0.1 and 0.2 pF. All of the circuit elements, except $Z$, are to be represented by a simpler Thevenin/Norton equivalent.

In this article, we present an empirical way of determining the Thevenin or Norton equivalent circuit corresponding to an actual bias circuit. This method assumes that the circuit does not change with time or temperature in the refrigerator. We also require that either the stray capacitance or stray inductance of the thermistor be negligible, so that either the current or voltage through the impedance $Z$ can be measured directly.

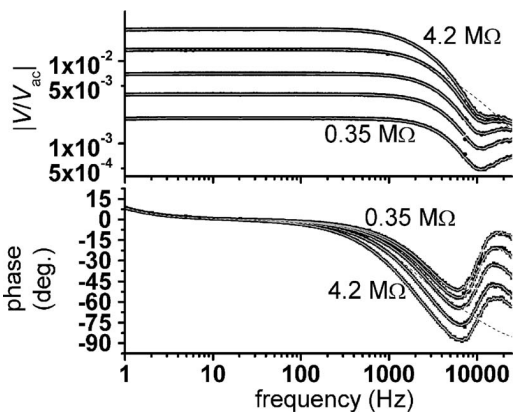

FIG. 2. Unbiased transfer functions. Each curve corresponds to a particular resistance of the thermistor. The resistances are 0.35, 0.68, 1.22, 2.40, 2.41, and 4.29 M$\Omega$. The black points represent the measured transfer function data. The white curves represent the fit of the Thevenin/Norton equivalent circuit model to the data. The equivalent circuit models the data accurately at all the measured frequencies. The dashed black curve represents the fit of the bias circuit model of Fig. 1 (with the additional high-pass and low pass filters included in the model). From this fit, we find the values of the capacitors in Fig. 1. The bias circuit model seems to fit the middle (2.4 M$\Omega$) transfer function for frequencies up to 1 kHz and diverges from the data at higher frequencies. However, the value of the capacitors that are yielded from the fit depends on which transfer function is used. At 10 Hz, the capacitances and electronic filters are negligible. We use this frequency to calculate the thermistor resistance. These data were measured with $V_{ac}$ ranging from 100 to 400 mV, but were scaled to correspond to a 100 mV bias.

## II. THEVENIN/NORTON EQUIVALENT CIRCUITS

Rather than construct an accurate physical model of the bias circuit, with its stray impedances, distributed loads, and the frequency dependent gains, we make use of Thevenin or Norton equivalence and at least two transfer functions to characterize the bias electronics. The bias circuit can be represented by a Thevenin equivalent voltage bias $V_{Th}$ in series with equivalent impedance $Z_{eq}$ and the thermistor impedance $Z$. Alternatively, the Norton equivalent circuit with the current source $I_N$ can represent the circuit. The Thevenin and Norton equivalents are related by $V_{Th} = I_N Z_{eq}$. In the following discussion, we use several unbiased transfer functions to solve for the circuit equivalents. In this discussion we will assume that the reactance of the unbiased thermistor remains fixed so that we can fully determine the equivalents from the transfer functions.

We will use measurements of our Si thermistors to illustrate the technique. When a thermistor is unbiased, the impedance $Z(f)$ is the resistance $R$, and is frequency independent. In each unbiased transfer function measurement, we measured the electronic response for hundreds of frequencies ranging from 1 Hz to 25 kHz. Between measurements, the resistance $R$ of the thermistor is varied by changing the temperature in the refrigerator. The temperature ranged from 60 to 150 mK. In Fig. 2, we plot a series of transfer functions corresponding to different thermistor resistances. In this plot, the gain of 923.6 has been divided out. Single pole high-pass and low-pass filters, at 0.156 Hz and 4.85 kHz, respectively, attenuate the measured data at lower and higher frequencies.

To compute the thermistor resistance, we only use data at a frequency such that both the bias circuit and readout electronics are well characterized. We choose a frequency $f_0$ (which is 10 Hz) at which the equivalent impedance of the bias circuit equals the load resistance ($R_L = 171.8$ M$\Omega$), and the Thevenin equivalent voltage $V_{Th}$ equals the ac bias voltage $V_{ac}$. At this frequency, the attenuation and phase shift of the electronic filters is negligible. The voltage $V(f_0)$ of across the thermistor resulting from bias $V_{ac}$ is found from the transfer functions. The resistance of the thermistor can be found from measurements of voltage $V(f_0)$ across the thermistor and Thevenin equivalent circuit by

$$R = Z_{eq}(f_0) \frac{V(f_0)}{V_{Th}(f_0) - V(f_0)}. \tag{1}$$

From the data in Fig. 2, we compute the following thermistor resistances: 0.35, 0.68, 1.22, 2.40, 2.41, and 4.29 M$\Omega$. We use the set of unbiased transfer functions and the associated measurements of resistance $R$ to calculate the Thevenin and Norton equivalents of the bias circuit for all measured frequencies.

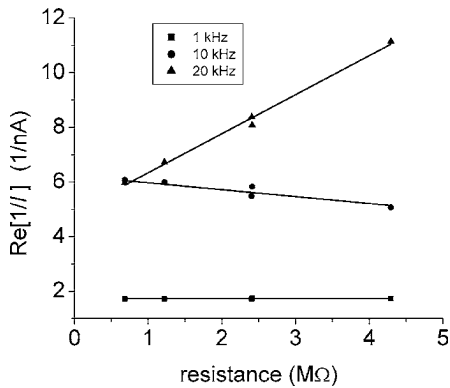

FIG. 3. Line fits to the inverse current. We plot the real part of the inverse current vs the resistance as determined from the unbiased transfer functions of Fig. 2. Thevenin voltages are found from the slope of the lines. The $x$ intercept and $y$ intercept determine the equivalent impedance and Norton current.

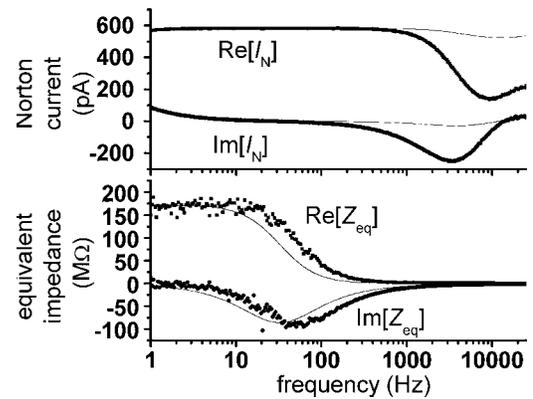

FIG. 4. The measured circuit equivalents. The plotted points represent measured values of the Norton current and the equivalent impedance. The thin curves represent equivalents of the circuit model in Fig. 1, with the electronic filters included. The curves diverge from the points because the best fit of that circuit does not accurately model the data at most frequencies. This demonstrates the usefulness of this approach.

For a particular ac voltage bias $V_{ac}$ applied to the circuit, the inverse of the ac current $I(f)$ is related to thermistor resistance $R$ by the following linear equation:

$$I^{-1} = V_{Th}^{-1} R + I_N^{-1}, \quad (2)$$

where $I$, $V_{Th}$, and $I_N$ are all complex functions of the frequency. This equation is true as long as the bias circuit and the bias voltage applied by the experimenter do not change between measurements. If the bias voltages are changed, then data from measurements at an alternative ac bias voltage $V_{ac2}$ should be normalized to the same ac bias voltage by multiplying $I$ by the ratio $(V_{ac2}/V_{ac})$. Fitting the inverse current versus resistance to the straight line of Eq. (2) provides a means to find the Thevenin and Norton equivalents of the bias circuit. The Thevenin voltage is found from the slope, and the Norton current is found from the $y$ intercept. To do this, we need two or more different unbiased transfer function measurements of the thermistor. Stray capacitance in the thermistor is treated as part of the bias circuit; it gets absorbed into the Thevenin and Norton equivalents.

Because $R$ is real, we may break up Eq. (2) into two real valued equations

$$\mathrm{Re}[y] = m_1 R + y_1 \quad (3)$$

and

$$\mathrm{Im}[y] = m_2 R + y_2, \quad (4)$$

where the slopes $m_1$ and $m_2$ are the real and imaginary parts of $(1/V_{Th})$ and $y_1$ and $y_2$ are the real and imaginary parts of $(1/I_N)$. We get a measurement of $y$ and $R$ for each frequency from the unbiased transfer functions. We find $m_1$, $m_2$, $y_1$, and $y_2$ by fitting the data to the two lines of Eqs. (3) and (4). Some examples of line fits to our unbiased transfer function data are shown in Fig. 3. Thevenin and Norton equivalents are found from the line fits using the following formulas:

$$V_{Th}(f) = (m_1 - i m_2)/(m_1^2 + m_2^2), \quad (5)$$

$$I_N(f) = (y_1 - i y_2)/(y_1^2 + y_2^2), \quad (6)$$

and

$$Z_{eq}(f) = V_{Th}/I_N. \quad (7)$$

The resulting values for the real and imaginary parts of $I_N$ and $Z_{eq}$ are plotted in Fig. 4. These data characterize the bias circuit. As a demonstration of the usefulness of this technique, we compare these equivalent circuit data to the circuit model of Fig. 1 (with the aforementioned high-pass and low-pass filters also included in the model). The circuit model was fit to a single unbiased transfer function in Fig. 2 in the manner described by Vaillancourt.[2] The Norton equivalent current and equivalent impedance corresponding to the best fit circuit model are plotted in Fig. 4 to compare to the measured data. The curves diverge from the data points because the circuit model does not accurately represent the real equivalent circuit data, even at frequencies less than 1 kHz. We could introduce additional free parameters to the model and fit to the equivalent circuit data, rather than a single transfer function. However, such modeling is not necessary in light of this approach because the measured circuit equivalents can be applied directly to find the impedance of a biased thermistor.

We now wish to use the Thevenin/Norton equivalents to compute the impedance of a biased thermistor, calorimeter, or bolometer. Note again that $V_{Th}$ and $I_N$ are proportional to the applied bias, so if the applied ac bias is different than $V_{ac}$, then $V_{Th}$ and $I_N$ should be scaled to correspond to bias $V_{ac}$. The thermistor impedance is obtained from transfer functions using the following transformation:

$$Z(f) = Z_{eq}(f) \frac{V(f)}{V_{Th}(f) - V(f)}. \quad (8)$$

If we apply this transformation to unbiased transfer functions we find $Z(f)$ equals the resistance $R$ of the thermistor for all measured frequencies, with some scatter associated with noise. Application of this transformation to a biased transfer function, such as is plotted in Fig. 5, results in $Z(f)$ of the biased calorimeter, shown in Fig. 6. In practice, we fit the calorimeter model to this impedance data to find calorimeter parameters such as heat capacities or thermal conductance.[1–6]

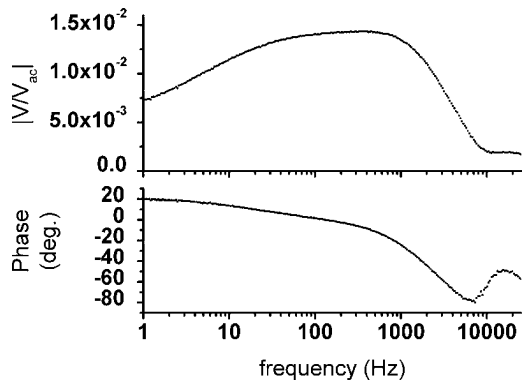

FIG. 5. The transfer function of a biased Si thermistor. The shape of this transfer function is different than the unbiased transfer functions of Fig. 2 due to electrothermal effects in the calorimeter.

We note that this transformation not only takes into account effects of the bias circuit and the stray capacitance in the thermistor, but other effects as well. Frequency dependent gains and phase shifts of the bias and readout electronics are also absorbed into $V_{Th}$ and $I_N$. As long as the electronics are stable throughout the measurements, we only need to know the electronic response near frequency $f_0$ to obtain accurate measurements of $V_{Th}$, $I_N$, and $Z(f)$ at any frequency in our range.

## III. CURRENT READOUT AND OTHER APPLICATIONS

In the above discussion, we treated data from the measured voltage across a Silicon thermistor. However, this technique is applicable to other kinds of devices in which the resistance of the device may be varied by the experimenter. For example, we have also applied the technique to transition-edge sensors (TESs), which are resistive thermometers consisting of a superconducting metal film on the phase transition. We read out the current through these devices with a Quantum Design dc SQUID. The resistance of a TES can be varied by either changing the magnetic field or the refrigerator temperature. The Norton current $I_N$ is directly obtained by measuring a transfer function when a TES is superconducting. Measurements of a second transfer function, with the TES in the normal state, allow the experimenter to obtain $V_{Th}$ and $Z_{eq}$. Inductance in a series with the TES gets absorbed into $Z_{eq}$. We then use the equivalent circuit parameters to obtain the impedance of the TES, when it is biased in the phase transition.

## IV. NOISE AND ERROR ESTIMATES

From the Thevenin equivalent circuit, impedance $Z$ is related to the voltage across a silicon thermistor by $Z/(Z+Z_{eq})=V/V_{Th}$. From this equation, we determine that the error $\Delta Z$ in the impedance $Z$ is related to noise $\Delta V$ in the thermistor voltage $V$ by

$$\Delta Z = \frac{dZ}{dV}\Delta V = \left| \frac{(Z_{eq}+Z)^2}{Z_{eq}} \frac{\Delta V}{V_{Th}} \right|. \quad (9)$$

Equation (9) gives the error due to noise in the biased transfer function data. Additional statistical error is associated with fitting linear equations [Eqs. (3) and (4)] to the unbiased transfer function data. To minimize that error, we measure unbiased transfer functions over a wide range of thermistor resistances ranging from minimum resistance $R_{min}$ up to maximum resistance $R_{max}$. The slope error of the linear fits causes an error in the calculated $Z$ that is proportional to $|Z-\bar{R}|$, where $\bar{R}$ is the average measured resistance. This error is small in the region on the complex plane defined by $|Z-\bar{R}|<R_{max}-R_{min}$. Within this circular region, the error in the calculated impedance $Z$ can be estimated using Eq. (9). Note that the error in the measurements of negative impedance $Z=-R$ (or an impedance with any nonzero phase) can be larger than the error associated with the measurement of the corresponding positive impedance $Z=+R$.

As an example, we estimate the noise in our measurements and compare it to the scatter in measured impedances in Fig. 6. At frequencies of less than 100 Hz we applied an ac bias $V_{ac}=2$ mV to our bias circuit in Fig. 1. For this bias circuit, $V_{Th}=V_{ac}$ and $Z_{eq}=200$ M$\Omega$ in this range of frequencies. The thermistor impedance $Z$ is 1.3 M$\Omega$. We measured 5 nV$/\sqrt{Hz}$ noise, as referred to the thermistor, and sampled it for 1 s. Therefore, we estimate the noise in our measurement of $Z$ to be about $\Delta Z=500$ $\Omega$, using Eq. (9). For our highest frequencies ($\sim$20 kHz) our bias electronics attenuate the ac bias so that $V_{ac}=0.4$ mV and $Z$ is 2.5 M$\Omega$. Stray capacitances in the bias circuit and a factor of 5 in attenuation in a readout preamp that gets absorbed into $V_{Th}$, reduce the Thevenin equivalents to $V_{Th}=800$ nV and $Z_{eq}=20$ k$\Omega$. At those frequencies our noise is estimated to be $\Delta Z=200$ k$\Omega$. Because the equivalent circuits are computed independently at each frequency and $Z(f)$ is a smooth function, we can use the scatter in Fig. 6 to measure the actual statistical error in the measurements. The actual scatter is in agreement with the above noise estimates. Besides noise, there are small additional errors associated with the determination of thermistor resistance from measurements at frequency $f_0$. These errors are about 1 part in a 1000.

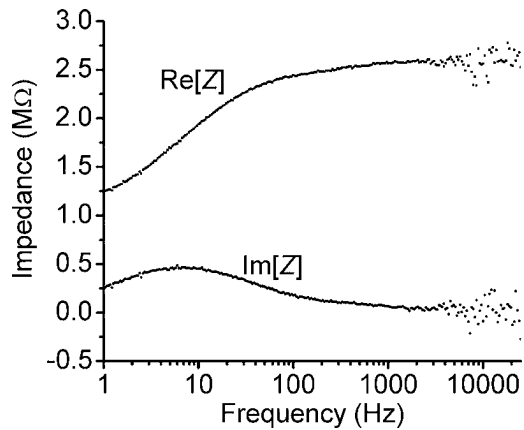

FIG. 6. The impedance of the biased calorimeter. These data are typical of what we expect from a calorimeter. If the real part is plotted vs the imaginary part, it traces a semicircular shape in the complex plane with some modification at high frequencies due to the physics of the coupling of the thermistor to the absorber of the calorimeter.

## V. DISCUSSION

The method described in this article provides an accurate instrument for measurements of impedance characteristics of calorimeter and bolometers. In our Si thermistor experiments, the quality of the resulting data is primarily limited by noise. In our data, the signal to noise ratio at a high frequency was somewhat degraded because our ac bias was significantly attenuated by our electronics at a high frequency. The quality of the data can be improved by increasing the ac bias or sampling longer at high frequencies.

The measured circuit equivalents can be used to construct a physical model of the bias electronics, which is useful for engineering the electronics. However, using such a model in conjunction with Eq. (8) to transform transfer functions into impedance characteristics could be detrimental. An incorrect assumption in the model could produce an artifact in the impedance data that could resemble structure in the impedance curve. Using equivalent circuit data that has been smoothed to minimize the noise could also generate such artifacts. Noise in the measured circuit equivalents has the advantage of being uncorrelated across frequencies. With this technique, any structure that is seen in the impedance data is likely to be real, as long as the bias circuit and electronics are stable throughout the measurements. Stability can be verified by repeat measurements of transfer functions with the same resistance and by verifying that data actually fit the straight lines of Eqs. (3) and (4) at all frequencies.